\documentclass[aps,english,nofootinbib,twocolumn, showkeys]{revtex4-1}
\usepackage[T1]{fontenc}
\usepackage[utf8]{inputenc}
\usepackage{xcolor}

\usepackage{babel}
\usepackage{pmboxdraw}
\usepackage{amstext}
\usepackage{graphicx}
\usepackage{bm}
\usepackage{amsmath}
\usepackage{makecell}
\usepackage{booktabs}
\usepackage[version=4]{mhchem}

\begin{document}


\title{Rapid Discovery of Graphene Nanocrystals Using DFT and Bayesian Optimization with Neural Network Kernel}
\author{\c{S}ener \"{O}z\"{o}nder}
\affiliation{Institute for Data Science \& Artificial Intelligence, Bo\u{g}azi\c{c}i University,
\.{I}stanbul, T\"{u}rkiye}
\email{Corresponding author: sener.ozonder@bogazici.edu.tr}
\author{H. K\"{u}bra K\"{u}\c{c}\"{u}kkartal}
\affiliation{Computer Engineering Department, Eski\c{s}ehir Osmangazi University, Eski\c{s}ehir, T\"{u}rkiye}
\affiliation{ArtificaX Technologies, Bo\u{g}azi\c{c}i Tecnopark, Sar{\i}yer \.{I}stanbul, T\"{u}rkiye}

\begin{abstract}

Density functional theory (DFT) is a powerful computational method used to obtain physical and chemical properties of materials. In the materials discovery framework, it is often necessary to virtually screen a large and high-dimensional chemical space to find materials with desired properties. However, grid searching a large chemical space with DFT is inefficient due to its high computational cost. We propose an approach utilizing Bayesian optimization (BO) with an artificial neural network kernel to enable smart search. This method leverages the BO algorithm, where the neural network, trained on a limited number of DFT results, determines the most promising regions of the chemical space to explore in subsequent iterations. This approach aims to discover materials with target properties while minimizing the number of DFT calculations required. To demonstrate the effectiveness of this method, we investigated 63 doped graphene quantum dots (GQDs) with sizes ranging from 1 to 2 nm to find the structure with the highest light absorbance. Using time-dependent DFT (TDDFT) only 12 times, we achieved a significant reduction in computational cost, approximately 20\% of what would be required for a full grid search, by employing the BO algorithm with a neural network kernel. Considering that TDDFT calculations for a single GQD require about half a day of wall time on high-performance computing nodes, this reduction is substantial. Our approach can be generalized to the discovery of new drugs, chemicals, crystals, and alloys with high-dimensional and large chemical spaces, offering a scalable solution for various applications in materials science.

\end{abstract}


\maketitle

\section{Introduction}

The computational and experimental studies of the properties of chemicals, crystals, drugs, biomaterials, and alloys require significant budgets and decades of dedicated work. Often, however, the outcomes are not practical for real-world applications. This materials-to-properties methodology can be revolutionized by new artificial intelligence and optimization methods. Instead of studying the properties of a single material and its structurally similar derivatives, we can search for the material with the target properties required for a specific application within the chemical and structural parameter space that contains all possible materials.

\begin{figure}[t!]
\centering
\includegraphics[scale=0.65]{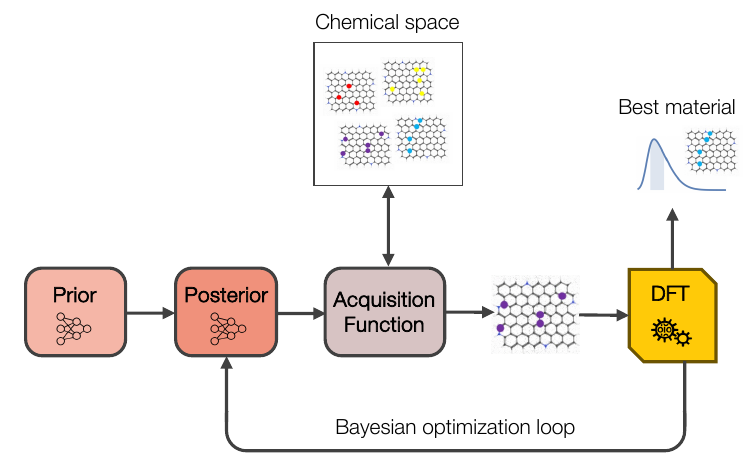}

\caption{Schematic of the Bayesian optimization loop for materials discovery. The process involves searching the chemical space of graphene quantum dots (GQDs) using density functional theory (DFT). The prior distribution is an artificial neural network (ANN) initialized with some past DFT results and the posterior distribution is the same ANN retrained with new results coming from the DFT evaluation in each iteration. The acquisition function selects the next point for DFT evaluation using the posterior, iteratively refining the model to identify the GQD with the highest absorption properties.}

\label{BO}
\end{figure}

In this computational materials discovery framework, the first step is to form the large and high-dimensional chemical space where each coordinate corresponds to a distinct structural, chemical or physical property \cite{Cheng2014, bera2014, Curtarolo2013, Paul2017, shen2022, hong2020, sharma2023, cai2020, bereau2016, sinnott2013, hautier2012, curtarolo2012, nandy2022}. These properties can be categorical or numerical (continuous or discrete). Secondly, the target property to be calculated computationally, such as absorption strength, conductivity, docking score, or another relevant metric, needs to be defined at every grid point of this chemical space. A smart search algorithm can then be used to model the whole chemical space. This algorithm shall invoke computationally costly simulations, such as density functional theory (DFT), molecular dynamics, or ligand docking, only a limited number of times. 
The surrogate model obtained in BO iterations is a probability distribution function of the target property over the whole chemical space and is therefore cheap and fast to call \cite{mak2017, sun2020, haghighat2021, wang2020, emiliano2017, antonello2023, chenglu2020, mcbride2019, anand2011, cozad2014, eason2014}. After a certain number of simulations (DFT or others), a faithful model of the chemical space is obtained, which can then be used iteratively to find the material in the whole chemical space that maximizes or minimizes the target property.

In this work, we propose an approach that searches the chemical space using Bayesian optimization (BO) with an artificial neural network (ANN) kernel (see Fig. \ref{BO}). We demonstrate the efficiency of this algorithm by testing it on a chemical space occupied by 63 distinct two-dimensional graphene quantum dots (GQD) with varying structural properties of size, dopant type, and dopant percentage. The target property is the total light absorbance in the 300-400 nm range, and the structure sought is the GQD that has the highest absorbance in that region of the spectrum. 
BO is an iterative method that uses the ANN as a surrogate model to determine the next batch of GQDs whose absorbance will be calculated via time-dependent density functional theory (TDDFT) \cite{wang2022, wangyifan2021, shields2021, florian2018, seongeon2018, agarwalgarvit2021, gao2022, fang2021, korovina2020, Ekstrom2019, motoyama2022, duris2020, roussel2021, Vargas2019}. This smart search algorithm avoids calling the costly DFT simulations for all 63 GQDs; only 12 DFT calculations were enough to find the GQD structure that has the maximum absorbance.

\begin{figure}[t!]
\includegraphics[width=.4\textwidth]{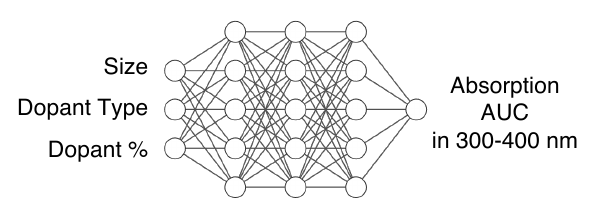}

\caption{Neural network architecture used as a kernel in the Bayesian optimization process. Inputs include GQD structural properties, and the target is the absorption area under the curve (AUC) in the 300-400 nm range. The acquisition function uses this neural network as a surrogate model to predict the target values for GQDs whose absorption AUC has yet to be calculated with DFT.}

\label{NN}
\end{figure}


The ``materials discovery'' paradigm aims to discover new materials that have the optimal chemical or physical properties best suited for the desired application \cite{oganov2019, zhanglijun2017, pan2018, wang2015, mcfarland1999, lookman2017, jansen2015, needs2016, bras2014, jain2016, roozbeh2017}. The goal is to rapidly discover the structure or chemical composition of the material that has the desired target property, such as the stoichiometric ratio giving rise to the alloy with the highest stiffness or the ligand having the right shape to dock tightly with a protein. Artificial intelligence approaches have recently proved to be useful for these types of tasks \cite{LeeByun,Zuo2021,Kusne2020, yueliu2017, yongfeijuan2021, raccuglia2016, saal2020, vasudevan2021, jihengfang2022, gubernatis2018, ioannis2024, quanzhou2018, peterson2021, tabor2018, merchant2023, yunxingzuo2021, song2021, lyngby2022}.


As a case study, we shall attempt to discover the GQD that absorbs the near UV-violet region (300-400 nm) most efficiently among others. High absorption in this region of the spectrum is useful for flame photodetectors, chemical sensors, bioimaging, solar cells, security inks and authentication devices \cite{ozonder-graphene-abs,koppens2014, yuxin2011, sarmento2018, chung2021, miao2012, mahmoudi2018, pang2018, hanyung2021}. Additionally, GQDs have interesting fluorescence properties where they can absorb near UV-violet light and emit visible light. This property is used in technological and scientific applications such as fluorescence microscopy, flow cytometry, and fluorescence-based sensors for detecting various biological and chemical substances \cite{liu2018, pirone2021, mitra2017, bingyang2021}. Hence, it is critical to computationally find the best absorbing GQD among all possible structures.


Considering that GQDs come in diverse shapes and dopant contents, the chemical space is vast, making it prohibitively expensive to use DFT for all possible GQDs. Therefore, a smart search algorithm is needed to explore the chemical space based on the gradients of the target property value. To have the target value change smoothly in this space, the coordinates must be chosen wisely. Here, we choose these coordinates to be graphene size, dopant type, and dopant percentage. We choose sizes of 1 nm, 1.5 nm, and 2 nm, and dopant percentages of 0\% (pristine), 1.5\%, 3\%, 5\%, and 7\%. For dopant types, we choose the elements B, N, O, S, and P. This creates a chemical space of $3\times5\times5$. Considering there are only three pristine GQDs, the number of distinct structures in this space is 63.

\begin{figure*}[!t]
\includegraphics[scale=0.62]{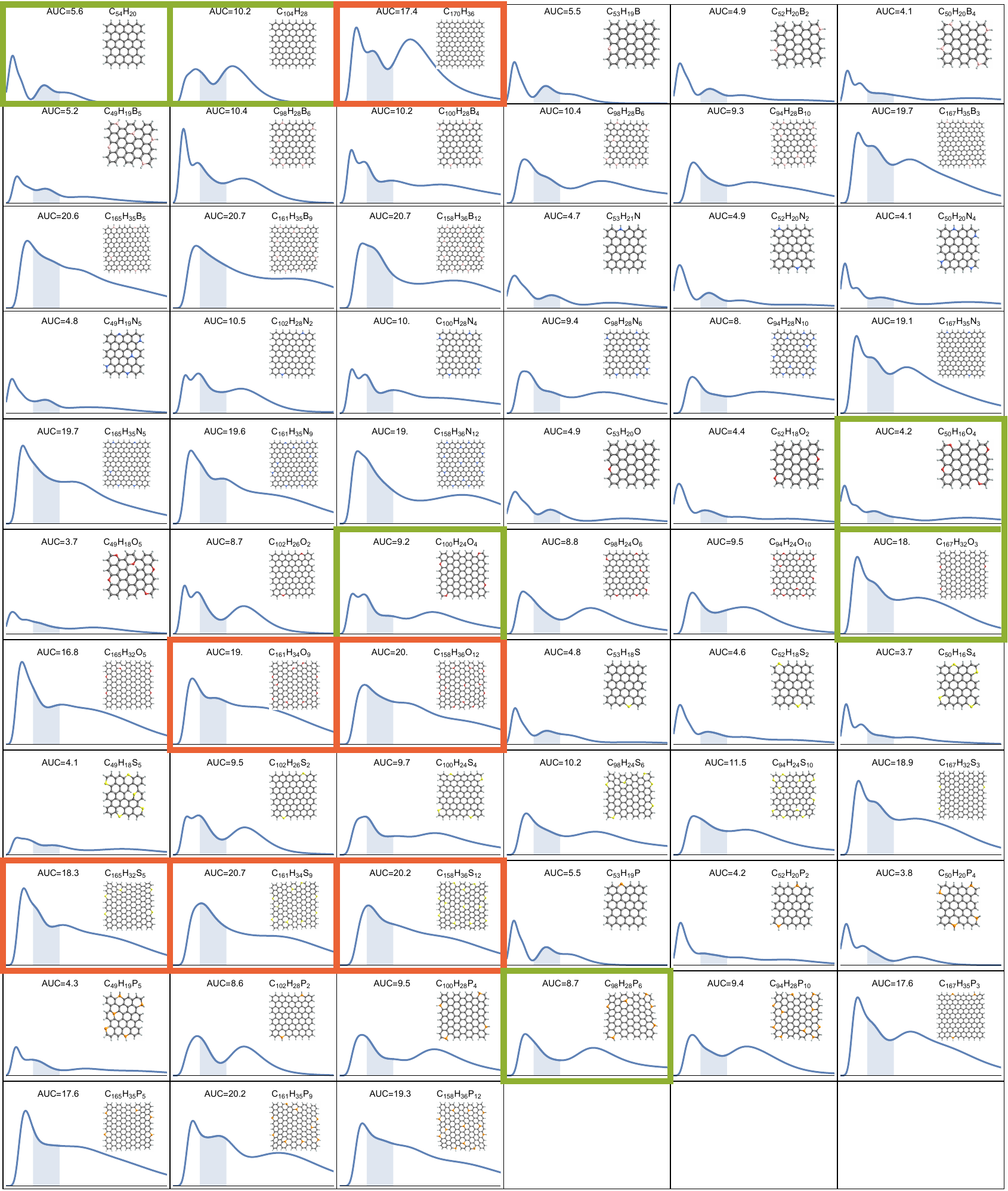}

\caption{All 63 GQD structures and their absorption curves calculated via TDDFT. The structures in green frames correspond to the initial dataset randomly chosen, whose DFT results were fed to the Bayesian optimization algorithm. The structures in orange frames are those selected by the acquisition function in each iteration. Bayesian optimization identifies only in six iterations one of the structures with the maximum absorption AUC (20.7) in the whole set, which is the 2 nm GQD S-doped at 5\% (\ce{C161H34S9}).} 

\label{structures}
\end{figure*}

Since dopant types have no obvious inherent order, unlike the numerical values of size and dopant percentage, it is not clear a priori how to order these elements on the dopant type coordinate. Here, we choose the ordering B, N, and O, which is the ordering in the second period of the periodic table. Next, we choose S since it is in the same group as the previous dopant element, O. In more complex scenarios where no guiding principle can be found for ordering categorical variables of a feature, either one-hot encoding can be used for each category, adopting each as a new coordinate that can assume values of 0 or 1, or the ordering of categorical variables can be determined using a learning algorithm such as a transformer with an attention matrix. The latter approach is used in drug discovery, where each element and character in the SMILES strings of ligands are taken as separate tokens with high-dimensional embedding vectors randomly initialized, and the embedding space is learned from the data via self-supervised learning \cite{griffithsrr2019, pinheiro2020, honda2019, aruspous2020, wangsheng2019}.


Once the chemical space is constructed, the surrogate model that captures the target value over the chemical space needs to be chosen. The surrogate model uses a limited number of past DFT results at some grid points in the chemical space and produces either a probability distribution function or directly the predicted value for the target over all grid points (GQD structures here) in the chemical space. A common choice is to use a Gaussian process to interpolate the target values from the past DFT results to grid points where DFT has yet to be used. However, Gaussian processes are known to be inefficient in capturing nonlinearities in high-dimensional data \cite{moriconi2020}. Furthermore, the kernel quantifying correlations between points must be hand-picked, not learned from the data. To avoid these problems, we use artificial neural networks where the inputs are size, dopant type, and dopant percentage, and the output is the target property. The network architecture includes three hidden layers with five neurons in each, containing ReLU activation functions (see Fig. \ref{NN}). Depending on the application, the target to be maximized can be the absorption peak value, area under the absorption curve (AUC) in a region of the spectrum, the number of peaks above a certain value, or the width of the absorption peak. In general, this target can be a scalar, vector, or even a tensor. In this work, we choose the target to be maximized as the AUC of the absorption curve in the 300-400 nm range.

\begin{figure*}[!t]
\includegraphics[scale=0.9]{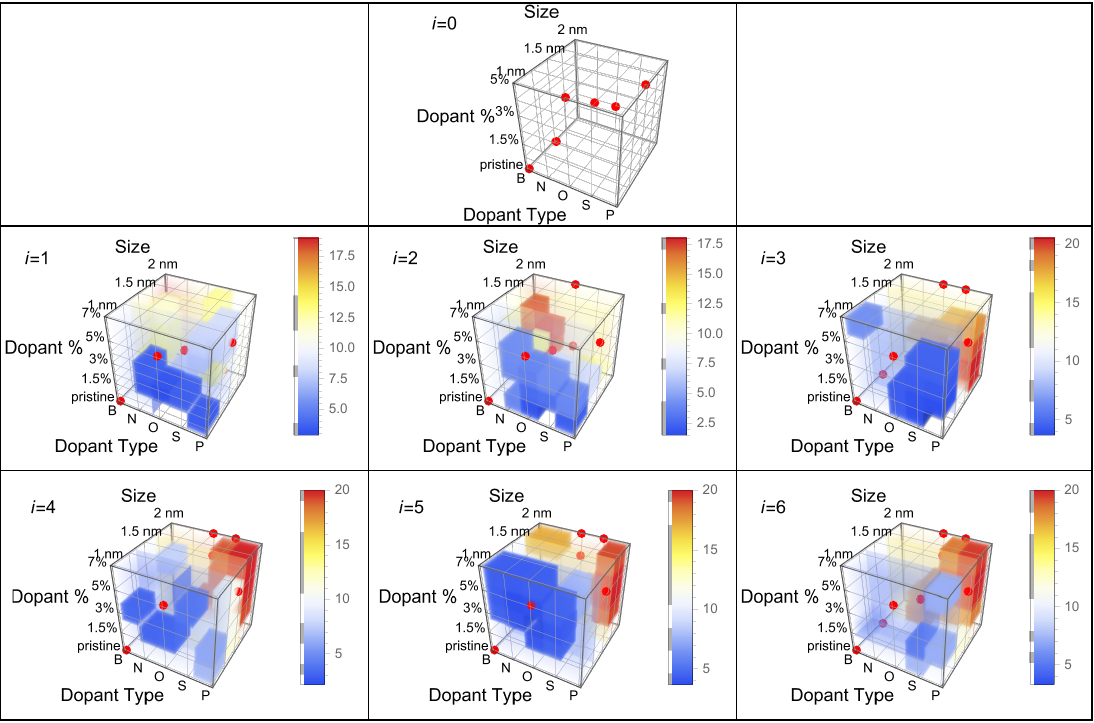}

\caption{The 3D density plot of neural network surrogate model for absorption AUC in the chemical search space. The evolution of the surrogate model after each iteration is depicted. As iterations continue, the density plot evolves, and the model becomes more confident. The colors in the bar legend signify the value of AUC, with red indicating the presence of GQDs with potentially high absorption AUC.}

\label{surrogate-sequences}
\end{figure*}


Having built the chemical space, defined the target property, and established the surrogate model, the BO algorithm can now be employed for material search. BO is an iterative method designed to find the maximum of an unknown objective function, $f(\bm{x})$, over the parameter space without evaluating $f(\bm{x})$ at all points within the parameter space $\bm{x} \in \chi$. The goal of BO is to determine the coordinates $\bm{x}^\star$ that maximize the objective function in the parameter space $\chi$.
\begin{equation} 
\bm{x}^\star = \underset{\bm{x} \in \chi}{\mathrm{argmax} }\, f(\bm{x})
\end{equation}

In each iteration, the acquisition function evaluates different points based on the posterior's mean and uncertainty to balance exploration (sampling where uncertainty is high) and exploitation (sampling where the mean prediction suggests high function values). The acquisition function's role is to identify the next point in the parameter space that is most promising for evaluation. This point is selected based on where the acquisition function indicates the highest potential for finding an optimal solution. After evaluating the objective function at this new point, the posterior distribution is updated with the new data, and the process repeats (see Fig.~\ref{BO}).

Application of BO to materials discovery has become popular in recent years \cite{packwood,LeeByun,ZUO2021126, WANG2022100728, Kusne2020, xus2021, pollice2021, doan2020, deshwal2021, shieldsb2021, WAHAB2020609, liang2021, GUO2023109714, terayama2021}. For example, it was used for automatic {\it de novo} molecular design in Ref. \cite{Bombarelli}. In another application, the minimum energy configuration for the grain boundary between Cu(001) and Cu(210) was found in 49 iterations among 17,983 possibilities \cite{Kiyohara}. This amounts to finding the target value by searching only 0.4\% of the whole parameter space. BO has also been to find the material with the smallest lattice thermal conductivity in 221 iterations among 54,799 candidate materials \cite{Seko2015}.

\section{Results and Discussion}

In this work, the objective function $f(\bm{x})$  is the target property of total absorbance (AUC) in the 300-400 nm region, and the parameter space $\chi$ is the $3\times5\times5$ chemical space defined by the coordinates: size, dopant type, and dopant percentage of GQDs. The algorithm begins by building the surrogate model through training the neural network with DFT results from six randomly selected GQDs. The acquisition function then suggests the next point in the search space (i.e., the next GQD structure) for which DFT will be conducted. This iterative process continues, with the neural network surrogate model being retrained with each new DFT result, thereby refining the posterior distribution and improving the search efficiency for the optimal GQD structure with maximum absorbance.

To initiate the BO algorithm, we calculated the absorption AUC of six randomly chosen GQDs via TDDFT. From the excited states, the absorption spectrum for the UV-Vis region was plotted, and the AUC for the 300-400 nm region was calculated (see the structures and their absorption curves in Fig. \ref{structures} in green frames).

The size, dopant type and dopant percentage of the six initial GQDs were used as input, and the AUC of each GQD was used as output to train a fully connected artificial neural network with ReLU activations. Next, an acquisition function using the probability of improvement method was employed along with this neural network surrogate model to determine the next GQD most likely to have the highest absorption AUC. DFT was then used to find the true AUC value of this seventh GQD. Now having one more data point regarding the true function, we retrained our neural network to build a better surrogate model, which in turn was used to determine the next GQD in the search. Six such iterations were conducted; the corresponding structures and their AUCs are shown in Fig. \ref{structures} in orange frames. It was found that the largest absorption AUC is 20.7, which belongs to the GQD of size 2 nm doped with S atoms at 5\% (see Table \ref{BOtable}).

To verify the results of the BO algorithm, we calculated the absorption AUC for all remaining 51 out of 63 GQDs using TDDFT. Among all 63 GQDs, three had the maximum absorption AUC of 20.7: 2 nm B-doped at 5\%, 2 nm B-doped at 7\%, and 2 nm S-doped at 5\%. Our BO experiment found the last one. This demonstrates that the absorption AUC is a target variable that changes smoothly enough with changing size and dopant coordinates in the search space, and BO can find one of the GQDs with the highest absorption AUC by calling DFT only on 12 GQDs out of 63. In a more extensive scenario where the search space includes additional coordinates such as solvent, surface functionals, shape, co-dopant type, and co-dopant percentage, and each coordinate has more resolution, the search space can grow from 63 GQDs to hundreds of thousands of GQDs. In this case, BO is expected to find the GQDs with target values that are at the maximum or in the vicinity of the maximum in the search space. This approach can be used along with any target property that is desired to be maximized depending on the application or research question.

The connection between the structural properties of GQDs and their target value (absorption AUC, in this case) is captured by the neural network surrogate model, as shown in Fig. \ref{surrogate-sequences}. The model estimates the absorption AUC for the grid points where DFT has not been used yet. The initial surrogate model is based on the randomly chosen six initial GQD absorption AUC values. With each additional DFT output introduced in each BO iteration, the model's confidence increases, which in turn guides the BO algorithm in determining which parts of the search space to explore or exploit next.

\begin{table}[h!]
\vspace{1cm}
\centering
\begin{tabular}{ccc}
\hline
\makecell{\textbf{Bayesian}\\ \textbf{optimization} \\ \textbf{iterations}} & \makecell{ \textbf{Graphene}\\ \textbf{structure}} &  \makecell{ \textbf{Absorption area}  \\ \textbf{(AUC) $( \times 10^{-6})$} }  \\
\hline
0 & 1 nm pristine & 5.6 \\
0 & 1.5 nm 5\% P-doped & 8.7 \\
0 & 1.5 nm 3\% O-doped & 9.2 \\
0 & 1.5 nm pristine & 10.2 \\
0 & 2 nm 1.5\% O-doped & 18.0 \\
0 & 1 nm \%5 O-doped & 4.2 \\
\hline
1 & 2 nm pristine & 17.4 \\
2 & 2 nm 7\% O-doped & 20.0 \\
3 & 2 nm 7\% S-doped & 20.2 \\
4 & 2 nm 5\% O-doped & 19 \\
\textbf{5} & \textbf{2 nm 5\% S-doped} & \textbf{20.7} \\
6 & 2 nm 3\% S-doped & 18.3 \\

\hline
\end{tabular}
\caption{Six initial GQDs and six iterations of Bayesian optimization, where absorption AUC is calculated from absorption curves derived via TDDFT. Bayesian optimization finds the structure with the maximum absorption AUC at the 5th iteration.}
\label{BOtable}
\end{table}

In summary, we presented a novel approach for the rapid discovery of graphene nanocrystals using density functional theory (DFT) and Bayesian optimization (BO) with neural network kernels. Our methodology efficiently navigates the chemical space to identify the graphene quantum dots (GQDs) with optimal absorption properties in the near UV-violet region (300-400 nm), significantly reducing the computational cost associated with exhaustive DFT evaluations. This methodology can be extended to other materials discovery applications, such as the optimization of electronic, thermal, and mechanical properties of nanomaterials, paving the way for significant advancements in materials science and engineering.

\section{Methods}
{\bf Description of dataset}

The input files for the 63 doped GQDs were prepared in Avogadro in XYZ format. All structures were set to the singlet state with zero charge. To ensure accurate bond counts, hydrogen atoms were added or removed as necessary.

{\bf TDDFT calculation details}

The structures were first optimized using the hybrid functional B3LYP with the basis set 6-31G(d) in Gaussian16. Simulations that resulted in imaginary frequencies were rerun with slightly different atomic positions. After geometry optimization, the excited states leading to the absorption spectrum were calculated via TDDFT using the same model chemistry. For both optimization and excited state calculations, water was used as the solvent in the polarizable continuum model (PCM) approach. 

Absorption plots are made by combining individual absorption curves of $n$ excited states via $\sum^n_{i=1} \varepsilon_i (\tilde{\nu})$. For each excited state $i$, we have
\begin{equation}
\varepsilon_i (\tilde{\nu}) = 1.3062974\times 10^8 \, \frac{f_i}{\sigma} \exp \left[ - \left( \frac{\tilde{\nu} -\tilde{\nu}_i}{\sigma}  \right)^2 \right]
\end{equation}
where $\varepsilon_i (\tilde{\nu})$ is in units of L mol$^{-1}$ cm$^{-1}$, $f_i$ is the oscillator strength of the $i$'th excited state and $\sigma=0.4$ eV. Here $\tilde{\nu}_i$ is the excitation energy (in wavenumbers), which can be related to wavelength in units of nm via $\tilde{\nu}_i = 1/\lambda_i$.

{\bf Acquisition function details} 

For the BO algorithm, Probability of Improvement (PI) approach was used as the acquisition function to balance exploration and exploitation. The PI at a point \( \bm{x} \) is given by:
\begin{equation}
\text{PI}(\bm{x}) = \Phi\left( \frac{\mu(\bm{x}) - f_{\text{best}} - \xi}{\sigma(\bm{x})} \right)
\end{equation}
where
 $\mu(\bm{x})$ is the predicted mean at point $\bm{x}$, 
$\sigma(\bm{x})$ is the predicted standard deviation at point $ \bm{x} $. Here $f_{\text{best}} $ is the best objective function value observed so far, $ \xi $ is an optional parameter that controls the trade-off between exploration and exploitation (in this case, 0.01).
$ \Phi $ is the cumulative distribution function of the standard normal distribution. 

This function quantifies the likelihood that sampling at \( \bm{x} \) will yield an improvement over the current best observation by at least \( \xi \). By selecting points that maximize this probability, the algorithm efficiently navigates the search space, balancing between exploring new regions and exploiting known promising areas.

{\bf Machine learning details}

The surrogate model that acted as the kernel of the BO algorithm was a fully connected feed forward network architecture with three hidden layers and five neurons in each layer, containing ReLU activation functions. ADAM was chosen as an optimizer and 1\% dropout was employed during training as a regularizer.

\section{Data Availability}
The DFT output files will be available from the corresponding author upon  request.

\section{Code Availability}
The code developed here will be available from the corresponding author upon request.

\section{Acknowledgement}
\c{S}.Ö. appreciates fruitful discussions with Mustafa Mısır on Bayesian optimization.
\c{S}.Ö. is supported
by TÜB\.{I}TAK under grant no. 120F354. Computing resources used in this
work were provided by the National Center for High Performance Computing
of T\"{u}rkiye (UHeM) under grant no. 1007872020 and TUBITAK ULAKBIM, High Performance and Grid Computing Center (TRUBA).

\bibliographystyle{naturemag}
\bibliography{BayesOptAbsorption}

\begin{thebibliography}{100}
\expandafter\ifx\csname url\endcsname\relax
  \def\url#1{\texttt{#1}}\fi
\expandafter\ifx\csname urlprefix\endcsname\relax\def\urlprefix{URL }\fi
\providecommand{\bibinfo}[2]{#2}
\providecommand{\eprint}[2][]{\url{#2}}

\bibitem{Cheng2014}
\bibinfo{author}{Cheng, X.}, \bibinfo{author}{Chen, X.-Q.},
  \bibinfo{author}{Li, D.-Z.} \& \bibinfo{author}{Li, Y.}
\newblock \bibinfo{title}{Computational materials discovery: the case of the
  w–b system}.
\newblock \emph{\bibinfo{journal}{Acta Crystallographica Section C}}
  \textbf{\bibinfo{volume}{70}}, \bibinfo{pages}{85--103}
  (\bibinfo{year}{2014}).

\bibitem{bera2014}
\bibinfo{author}{Bera, C.} \emph{et~al.}
\newblock \bibinfo{title}{Integrated computational materials discovery of
  silver doped tin sulfide as a thermoelectric material}.
\newblock \emph{\bibinfo{journal}{Phys. Chem. Chem. Phys.}}
  \textbf{\bibinfo{volume}{16}}, \bibinfo{pages}{19894--19899}
  (\bibinfo{year}{2014}).

\bibitem{Curtarolo2013}
\bibinfo{author}{Curtarolo, S.} \emph{et~al.}
\newblock \bibinfo{title}{The high-throughput highway to computational
  materials design}.
\newblock \emph{\bibinfo{journal}{Nature materials}}
  \textbf{\bibinfo{volume}{12}}, \bibinfo{pages}{191--201}
  (\bibinfo{year}{2013}).

\bibitem{Paul2017}
\bibinfo{author}{Paul, J.} \emph{et~al.}
\newblock \bibinfo{title}{Computational methods for 2d materials: Discovery,
  property characterization, and application design}.
\newblock \emph{\bibinfo{journal}{Journal of Physics: Condensed Matter}}
  \textbf{\bibinfo{volume}{29}} (\bibinfo{year}{2017}).

\bibitem{shen2022}
\bibinfo{author}{Shen, L.}, \bibinfo{author}{Zhou, J.}, \bibinfo{author}{Yang,
  T.}, \bibinfo{author}{Yang, M.} \& \bibinfo{author}{Feng, Y.~P.}
\newblock \bibinfo{title}{High-throughput computational discovery and
  intelligent design of two-dimensional functional materials for various
  applications}.
\newblock \emph{\bibinfo{journal}{Accounts of Materials Research}}
  \textbf{\bibinfo{volume}{3}}, \bibinfo{pages}{572--583}
  (\bibinfo{year}{2022}).

\bibitem{hong2020}
\bibinfo{author}{Hong, Y.}, \bibinfo{author}{Hou, B.}, \bibinfo{author}{Jiang,
  H.} \& \bibinfo{author}{Zhang, J.}
\newblock \bibinfo{title}{Machine learning and artificial neural network
  accelerated computational discoveries in materials science}.
\newblock \emph{\bibinfo{journal}{WIREs Computational Molecular Science}}
  \textbf{\bibinfo{volume}{10}}, \bibinfo{pages}{e1450}.

\bibitem{sharma2023}
\bibinfo{author}{Sharma, H.} \emph{et~al.}
\newblock \bibinfo{title}{Computational materials discovery and development for
  li and non-li advanced battery chemistries: Review paper}.
\newblock \emph{\bibinfo{journal}{Journal of Electrochemical Science and
  Engineering}}  (\bibinfo{year}{2023}).

\bibitem{cai2020}
\bibinfo{author}{Cai, J.}, \bibinfo{author}{Chu, X.}, \bibinfo{author}{Xu, K.},
  \bibinfo{author}{Li, H.} \& \bibinfo{author}{Wei, J.}
\newblock \bibinfo{title}{Machine learning-driven new material discovery}.
\newblock \emph{\bibinfo{journal}{Nanoscale Adv.}}
  \textbf{\bibinfo{volume}{2}}, \bibinfo{pages}{3115--3130}
  (\bibinfo{year}{2020}).

\bibitem{bereau2016}
\bibinfo{author}{Bereau, T.}, \bibinfo{author}{Andrienko, D.} \&
  \bibinfo{author}{Kremer, K.}
\newblock \bibinfo{title}{{Research Update: Computational materials discovery
  in soft matter}}.
\newblock \emph{\bibinfo{journal}{APL Materials}} \textbf{\bibinfo{volume}{4}},
  \bibinfo{pages}{053101} (\bibinfo{year}{2016}).

\bibitem{sinnott2013}
\bibinfo{author}{Sinnott, S.~B.}
\newblock \bibinfo{title}{{Material design and discovery with computational
  materials science}}.
\newblock \emph{\bibinfo{journal}{Journal of Vacuum Science Technology A}}
  \textbf{\bibinfo{volume}{31}}, \bibinfo{pages}{050812}
  (\bibinfo{year}{2013}).

\bibitem{hautier2012}
\bibinfo{author}{Hautier, G.}, \bibinfo{author}{Jain, A.},
  \bibinfo{author}{Shyue, b.} \& \bibinfo{author}{Ong, S.}
\newblock \bibinfo{title}{From the computer to the laboratory: Materials
  discovery and design using first-principles calculations}.
\newblock \emph{\bibinfo{journal}{Journal of Materials Science}}
  \textbf{\bibinfo{volume}{47}} (\bibinfo{year}{2012}).

\bibitem{curtarolo2012}
\bibinfo{author}{Curtarolo, S.} \emph{et~al.}
\newblock \bibinfo{title}{Aflow: An automatic framework for high-throughput
  materials discovery}.
\newblock \emph{\bibinfo{journal}{Computational Materials Science}}
  \textbf{\bibinfo{volume}{58}}, \bibinfo{pages}{218--226}
  (\bibinfo{year}{2012}).

\bibitem{nandy2022}
\bibinfo{author}{Nandy, A.}, \bibinfo{author}{Duan, C.} \&
  \bibinfo{author}{Kulik, H.~J.}
\newblock \bibinfo{title}{Audacity of huge: overcoming challenges of data
  scarcity and data quality for machine learning in computational materials
  discovery}.
\newblock \emph{\bibinfo{journal}{Current Opinion in Chemical Engineering}}
  \textbf{\bibinfo{volume}{36}}, \bibinfo{pages}{100778}
  (\bibinfo{year}{2022}).

\bibitem{mak2017}
\bibinfo{author}{Mak, S.} \emph{et~al.}
\newblock \bibinfo{title}{An efficient surrogate model for emulation and
  physics extraction of large eddy simulations}.
\newblock \emph{\bibinfo{journal}{Journal of the Acoustical Society of America
  (JASA)}} \textbf{\bibinfo{volume}{113}} (\bibinfo{year}{2017}).

\bibitem{sun2020}
\bibinfo{author}{Sun, L.}, \bibinfo{author}{Gao, H.}, \bibinfo{author}{Pan, S.}
  \& \bibinfo{author}{Wang, J.-X.}
\newblock \bibinfo{title}{Surrogate modeling for fluid flows based on
  physics-constrained deep learning without simulation data}.
\newblock \emph{\bibinfo{journal}{Computer Methods in Applied Mechanics and
  Engineering}} \textbf{\bibinfo{volume}{361}}, \bibinfo{pages}{112732}
  (\bibinfo{year}{2020}).

\bibitem{haghighat2021}
\bibinfo{author}{Haghighat, E.}, \bibinfo{author}{Raissi, M.},
  \bibinfo{author}{Moure, A.}, \bibinfo{author}{Gomez, H.} \&
  \bibinfo{author}{Juanes, R.}
\newblock \bibinfo{title}{A physics-informed deep learning framework for
  inversion and surrogate modeling in solid mechanics}.
\newblock \emph{\bibinfo{journal}{Computer Methods in Applied Mechanics and
  Engineering}} \textbf{\bibinfo{volume}{379}}, \bibinfo{pages}{113741}
  (\bibinfo{year}{2021}).

\bibitem{wang2020}
\bibinfo{author}{Wang, B.}, \bibinfo{author}{Zhang, G.}, \bibinfo{author}{Wang,
  H.}, \bibinfo{author}{Xuan, J.} \& \bibinfo{author}{Jiao, K.}
\newblock \bibinfo{title}{Multi-physics-resolved digital twin of proton
  exchange membrane fuel cells with a data-driven surrogate model}.
\newblock \emph{\bibinfo{journal}{Energy and AI}} \textbf{\bibinfo{volume}{1}},
  \bibinfo{pages}{100004} (\bibinfo{year}{2020}).

\bibitem{emiliano2017}
\bibinfo{author}{Iuliano, E.}
\newblock \bibinfo{title}{Global optimization of benchmark aerodynamic cases
  using physics-based surrogate models}.
\newblock \emph{\bibinfo{journal}{Aerospace Science and Technology}}
  \textbf{\bibinfo{volume}{67}}, \bibinfo{pages}{273--286}
  (\bibinfo{year}{2017}).

\bibitem{antonello2023}
\bibinfo{author}{Antonello, F.}, \bibinfo{author}{Buongiorno, J.} \&
  \bibinfo{author}{Zio, E.}
\newblock \bibinfo{title}{Physics informed neural networks for surrogate
  modeling of accidental scenarios in nuclear power plants}.
\newblock \emph{\bibinfo{journal}{Nuclear Engineering and Technology}}
  \textbf{\bibinfo{volume}{55}}, \bibinfo{pages}{3409--3416}
  (\bibinfo{year}{2023}).

\bibitem{chenglu2020}
\bibinfo{author}{Lu, C.}, \bibinfo{author}{Fei, C.-W.}, \bibinfo{author}{Liu,
  H.-T.}, \bibinfo{author}{Li, H.} \& \bibinfo{author}{An, L.-Q.}
\newblock \bibinfo{title}{Moving extremum surrogate modeling strategy for
  dynamic reliability estimation of turbine blisk with multi-physics fields}.
\newblock \emph{\bibinfo{journal}{Aerospace Science and Technology}}
  \textbf{\bibinfo{volume}{106}}, \bibinfo{pages}{106112}
  (\bibinfo{year}{2020}).

\bibitem{mcbride2019}
\bibinfo{author}{McBride, K.} \& \bibinfo{author}{Sundmacher, K.}
\newblock \bibinfo{title}{Overview of surrogate modeling in chemical process
  engineering}.
\newblock \emph{\bibinfo{journal}{Chemie Ingenieur Technik}}
  \textbf{\bibinfo{volume}{91}}, \bibinfo{pages}{228--239}
  (\bibinfo{year}{2019}).

\bibitem{anand2011}
\bibinfo{author}{Anand, K.}, \bibinfo{author}{Ra, Y.}, \bibinfo{author}{Reitz,
  R.~D.} \& \bibinfo{author}{Bunting, B.}
\newblock \bibinfo{title}{Surrogate model development for fuels for advanced
  combustion engines}.
\newblock \emph{\bibinfo{journal}{Energy \& Fuels}}
  \textbf{\bibinfo{volume}{25}}, \bibinfo{pages}{1474--1484}
  (\bibinfo{year}{2011}).

\bibitem{cozad2014}
\bibinfo{author}{Cozad, A.}, \bibinfo{author}{Sahinidis, N.~V.} \&
  \bibinfo{author}{Miller, D.~C.}
\newblock \bibinfo{title}{Learning surrogate models for simulation-based
  optimization}.
\newblock \emph{\bibinfo{journal}{AIChE Journal}}
  \textbf{\bibinfo{volume}{60}}, \bibinfo{pages}{2211--2227}
  (\bibinfo{year}{2014}).

\bibitem{eason2014}
\bibinfo{author}{Eason, J.} \& \bibinfo{author}{Cremaschi, S.}
\newblock \bibinfo{title}{Adaptive sequential sampling for surrogate model
  generation with artificial neural networks}.
\newblock \emph{\bibinfo{journal}{Computers \& Chemical Engineering}}
  \textbf{\bibinfo{volume}{68}}, \bibinfo{pages}{220--232}
  (\bibinfo{year}{2014}).

\bibitem{wang2022}
\bibinfo{author}{Wang, K.} \& \bibinfo{author}{Dowling, A.~W.}
\newblock \bibinfo{title}{Bayesian optimization for chemical products and
  functional materials}.
\newblock \emph{\bibinfo{journal}{Current Opinion in Chemical Engineering}}
  \textbf{\bibinfo{volume}{36}}, \bibinfo{pages}{100728}
  (\bibinfo{year}{2022}).

\bibitem{wangyifan2021}
\bibinfo{author}{Wang, Y.}, \bibinfo{author}{Chen, T.-Y.} \&
  \bibinfo{author}{Vlachos, D.~G.}
\newblock \bibinfo{title}{Nextorch: A design and bayesian optimization toolkit
  for chemical sciences and engineering}.
\newblock \emph{\bibinfo{journal}{Journal of Chemical Information and
  Modeling}} \textbf{\bibinfo{volume}{61}}, \bibinfo{pages}{5312--5319}
  (\bibinfo{year}{2021}).

\bibitem{shields2021}
\bibinfo{author}{Shields, B.} \emph{et~al.}
\newblock \bibinfo{title}{Bayesian reaction optimization as a tool for chemical
  synthesis}.
\newblock \emph{\bibinfo{journal}{Nature}} \textbf{\bibinfo{volume}{590}},
  \bibinfo{pages}{89--96} (\bibinfo{year}{2021}).

\bibitem{florian2018}
\bibinfo{author}{Häse, F.}, \bibinfo{author}{Roch, L.~M.},
  \bibinfo{author}{Kreisbeck, C.} \& \bibinfo{author}{Aspuru-Guzik, A.}
\newblock \bibinfo{title}{Phoenics: A bayesian optimizer for chemistry}.
\newblock \emph{\bibinfo{journal}{ACS Central Science}}
  \textbf{\bibinfo{volume}{4}}, \bibinfo{pages}{1134--1145}
  (\bibinfo{year}{2018}).

\bibitem{seongeon2018}
\bibinfo{author}{Park, S.}, \bibinfo{author}{Na, J.}, \bibinfo{author}{Kim, M.}
  \& \bibinfo{author}{Lee, J.~M.}
\newblock \bibinfo{title}{Multi-objective bayesian optimization of chemical
  reactor design using computational fluid dynamics}.
\newblock \emph{\bibinfo{journal}{Computers \& Chemical Engineering}}
  \textbf{\bibinfo{volume}{119}}, \bibinfo{pages}{25--37}
  (\bibinfo{year}{2018}).

\bibitem{agarwalgarvit2021}
\bibinfo{author}{Agarwal, G.}, \bibinfo{author}{Doan, H.~A.},
  \bibinfo{author}{Robertson, L.~A.}, \bibinfo{author}{Zhang, L.} \&
  \bibinfo{author}{Assary, R.~S.}
\newblock \bibinfo{title}{Discovery of energy storage molecular materials using
  quantum chemistry-guided multiobjective bayesian optimization}.
\newblock \emph{\bibinfo{journal}{Chemistry of Materials}}
  \textbf{\bibinfo{volume}{33}}, \bibinfo{pages}{8133--8144}
  (\bibinfo{year}{2021}).

\bibitem{gao2022}
\bibinfo{author}{Gao, H.} \emph{et~al.}
\newblock \bibinfo{title}{Revolutionizing membrane design using machine
  learning-bayesian optimization}.
\newblock \emph{\bibinfo{journal}{Environmental Science \& Technology}}
  \textbf{\bibinfo{volume}{56}}, \bibinfo{pages}{2572--2581}
  (\bibinfo{year}{2022}).

\bibitem{fang2021}
\bibinfo{author}{Fang, L.}, \bibinfo{author}{Makkonen, E.},
  \bibinfo{author}{Todorović, M.}, \bibinfo{author}{Rinke, P.} \&
  \bibinfo{author}{Chen, X.}
\newblock \bibinfo{title}{Efficient amino acid conformer search with bayesian
  optimization}.
\newblock \emph{\bibinfo{journal}{Journal of Chemical Theory and Computation}}
  \textbf{\bibinfo{volume}{17}}, \bibinfo{pages}{1955--1966}
  (\bibinfo{year}{2021}).

\bibitem{korovina2020}
\bibinfo{author}{Korovina, K.} \emph{et~al.}
\newblock \bibinfo{title}{Chembo: Bayesian optimization of small organic
  molecules with synthesizable recommendations}.
\newblock In \bibinfo{editor}{Chiappa, S.} \& \bibinfo{editor}{Calandra, R.}
  (eds.) \emph{\bibinfo{booktitle}{Proceedings of the Twenty Third
  International Conference on Artificial Intelligence and Statistics}}, vol.
  \bibinfo{volume}{108} of \emph{\bibinfo{series}{Proceedings of Machine
  Learning Research}}, \bibinfo{pages}{3393--3403} (\bibinfo{publisher}{PMLR},
  \bibinfo{year}{2020}).

\bibitem{Ekstrom2019}
\bibinfo{author}{Ekström, A.} \emph{et~al.}
\newblock \bibinfo{title}{Bayesian optimization in ab initio nuclear physics}.
\newblock \emph{\bibinfo{journal}{Journal of Physics G: Nuclear and Particle
  Physics}} \textbf{\bibinfo{volume}{46}}, \bibinfo{pages}{095101}
  (\bibinfo{year}{2019}).

\bibitem{motoyama2022}
\bibinfo{author}{Motoyama, Y.} \emph{et~al.}
\newblock \bibinfo{title}{Bayesian optimization package: Physbo}.
\newblock \emph{\bibinfo{journal}{Computer Physics Communications}}
  \textbf{\bibinfo{volume}{278}}, \bibinfo{pages}{108405}
  (\bibinfo{year}{2022}).

\bibitem{duris2020}
\bibinfo{author}{Duris, J.} \emph{et~al.}
\newblock \bibinfo{title}{Bayesian optimization of a free-electron laser}.
\newblock \emph{\bibinfo{journal}{Phys. Rev. Lett.}}
  \textbf{\bibinfo{volume}{124}}, \bibinfo{pages}{124801}
  (\bibinfo{year}{2020}).

\bibitem{roussel2021}
\bibinfo{author}{Roussel, R.}, \bibinfo{author}{Hanuka, A.} \&
  \bibinfo{author}{Edelen, A.}
\newblock \bibinfo{title}{Multiobjective bayesian optimization for online
  accelerator tuning}.
\newblock \emph{\bibinfo{journal}{Phys. Rev. Accel. Beams}}
  \textbf{\bibinfo{volume}{24}}, \bibinfo{pages}{062801}
  (\bibinfo{year}{2021}).

\bibitem{Vargas2019}
\bibinfo{author}{Vargas-Hernández, R.~A.}, \bibinfo{author}{Guan, Y.},
  \bibinfo{author}{Zhang, D.~H.} \& \bibinfo{author}{Krems, R.~V.}
\newblock \bibinfo{title}{Bayesian optimization for the inverse scattering
  problem in quantum reaction dynamics}.
\newblock \emph{\bibinfo{journal}{New Journal of Physics}}
  \textbf{\bibinfo{volume}{21}}, \bibinfo{pages}{022001}
  (\bibinfo{year}{2019}).

\bibitem{oganov2019}
\bibinfo{author}{Oganov, A.}, \bibinfo{author}{Pickard, C.},
  \bibinfo{author}{Zhu, Q.} \& \bibinfo{author}{Needs, R.}
\newblock \bibinfo{title}{Structure prediction drives materials discovery}.
\newblock \emph{\bibinfo{journal}{Nature Reviews Materials}}
  \textbf{\bibinfo{volume}{4}} (\bibinfo{year}{2019}).

\bibitem{zhanglijun2017}
\bibinfo{author}{Zhang, L.}, \bibinfo{author}{Wang, Y.}, \bibinfo{author}{Lv,
  J.} \& \bibinfo{author}{ma, Y.}
\newblock \bibinfo{title}{Materials discovery at high pressures}.
\newblock \emph{\bibinfo{journal}{Nature Reviews Materials}}
  \textbf{\bibinfo{volume}{2}}, \bibinfo{pages}{17005} (\bibinfo{year}{2017}).

\bibitem{pan2018}
\bibinfo{author}{Pan, J.} \& \bibinfo{author}{Yan, Q.}
\newblock \bibinfo{title}{Data-driven material discovery for photocatalysis: a
  short review}.
\newblock \emph{\bibinfo{journal}{Journal of Semiconductors}}
  \textbf{\bibinfo{volume}{39}}, \bibinfo{pages}{071001}
  (\bibinfo{year}{2018}).

\bibitem{wang2015}
\bibinfo{author}{Wang, Y.} \emph{et~al.}
\newblock \bibinfo{title}{Materials discovery via calypso methodology}.
\newblock \emph{\bibinfo{journal}{Journal of Physics: Condensed Matter}}
  \textbf{\bibinfo{volume}{27}}, \bibinfo{pages}{203203}
  (\bibinfo{year}{2015}).

\bibitem{mcfarland1999}
\bibinfo{author}{McFarland, E.~W.} \& \bibinfo{author}{Weinberg, W.}
\newblock \bibinfo{title}{Combinatorial approaches to materials discovery}.
\newblock \emph{\bibinfo{journal}{Trends in Biotechnology}}
  \textbf{\bibinfo{volume}{17}}, \bibinfo{pages}{107--115}
  (\bibinfo{year}{1999}).

\bibitem{lookman2017}
\bibinfo{author}{Lookman, T.}, \bibinfo{author}{Balachandran, P.~V.},
  \bibinfo{author}{Xue, D.}, \bibinfo{author}{Hogden, J.} \&
  \bibinfo{author}{Theiler, J.}
\newblock \bibinfo{title}{Statistical inference and adaptive design for
  materials discovery}.
\newblock \emph{\bibinfo{journal}{Current Opinion in Solid State and Materials
  Science}} \textbf{\bibinfo{volume}{21}}, \bibinfo{pages}{121--128}
  (\bibinfo{year}{2017}).

\bibitem{jansen2015}
\bibinfo{author}{Jansen, M.}
\newblock \bibinfo{title}{Conceptual inorganic materials discovery – a road
  map}.
\newblock \emph{\bibinfo{journal}{Advanced Materials}}
  \textbf{\bibinfo{volume}{27}}, \bibinfo{pages}{3229--3242}
  (\bibinfo{year}{2015}).

\bibitem{needs2016}
\bibinfo{author}{Needs, R.~J.} \& \bibinfo{author}{Pickard, C.~J.}
\newblock \bibinfo{title}{{Perspective: Role of structure prediction in
  materials discovery and design}}.
\newblock \emph{\bibinfo{journal}{APL Materials}} \textbf{\bibinfo{volume}{4}},
  \bibinfo{pages}{053210} (\bibinfo{year}{2016}).

\bibitem{bras2014}
\bibinfo{author}{Le~Bras, R.} \emph{et~al.}
\newblock \bibinfo{title}{A computational challenge problem in materials
  discovery: synthetic problem generator and real-world datasets}.
\newblock In \emph{\bibinfo{booktitle}{Proceedings of the Twenty-Eighth AAAI
  Conference on Artificial Intelligence}}, AAAI'14, \bibinfo{pages}{438–443}
  (\bibinfo{publisher}{AAAI Press}, \bibinfo{year}{2014}).

\bibitem{jain2016}
\bibinfo{author}{Jain, A.}, \bibinfo{author}{Shin, Y.} \&
  \bibinfo{author}{Persson, K.}
\newblock \bibinfo{title}{Computational predictions of energy materials using
  density functional theory}.
\newblock \emph{\bibinfo{journal}{Nature Review Materials}}
  \textbf{\bibinfo{volume}{1}}, \bibinfo{pages}{15004} (\bibinfo{year}{2016}).

\bibitem{roozbeh2017}
\bibinfo{author}{Dehghannasiri, R.} \emph{et~al.}
\newblock \bibinfo{title}{Optimal experimental design for materials discovery}.
\newblock \emph{\bibinfo{journal}{Computational Materials Science}}
  \textbf{\bibinfo{volume}{129}}, \bibinfo{pages}{311--322}
  (\bibinfo{year}{2017}).

\bibitem{LeeByun}
\bibinfo{author}{Lee, S.}, \bibinfo{author}{Byun, H.}, \bibinfo{author}{Cheon,
  M.}, \bibinfo{author}{Kim, J.} \& \bibinfo{author}{Lee, J.~H.}
\newblock \bibinfo{title}{Machine learning-based discovery of molecules,
  crystals, and composites: A perspective review}.
\newblock \emph{\bibinfo{journal}{Korean J. Chem. Eng.}}
  \textbf{\bibinfo{volume}{38}}, \bibinfo{pages}{1971--1982}
  (\bibinfo{year}{2021}).

\bibitem{Zuo2021}
\bibinfo{author}{Zuo, Y.} \emph{et~al.}
\newblock \bibinfo{title}{Accelerating materials discovery with bayesian
  optimization and graph deep learning}.
\newblock \emph{\bibinfo{journal}{Materials Today}}
  \textbf{\bibinfo{volume}{51}}, \bibinfo{pages}{126--135}
  (\bibinfo{year}{2021}).

\bibitem{Kusne2020}
\bibinfo{author}{Kusne, A.~G.} \emph{et~al.}
\newblock \bibinfo{title}{On-the-fly closed-loop materials discovery via
  bayesian active learning}.
\newblock \emph{\bibinfo{journal}{Nature Communications}}
  \textbf{\bibinfo{volume}{11}} (\bibinfo{year}{2020}).

\bibitem{yueliu2017}
\bibinfo{author}{Liu, Y.}, \bibinfo{author}{Zhao, T.}, \bibinfo{author}{Ju, W.}
  \& \bibinfo{author}{Shi, S.}
\newblock \bibinfo{title}{Materials discovery and design using machine
  learning}.
\newblock \emph{\bibinfo{journal}{Journal of Materiomics}}
  \textbf{\bibinfo{volume}{3}}, \bibinfo{pages}{159--177}
  (\bibinfo{year}{2017}).

\bibitem{yongfeijuan2021}
\bibinfo{author}{Juan, Y.}, \bibinfo{author}{Dai, Y.}, \bibinfo{author}{Yang,
  Y.} \& \bibinfo{author}{Zhang, J.}
\newblock \bibinfo{title}{Accelerating materials discovery using machine
  learning}.
\newblock \emph{\bibinfo{journal}{Journal of Materials Science \& Technology}}
  \textbf{\bibinfo{volume}{79}}, \bibinfo{pages}{178--190}
  (\bibinfo{year}{2021}).

\bibitem{raccuglia2016}
\bibinfo{author}{Raccuglia, P.} \emph{et~al.}
\newblock \bibinfo{title}{Machine-learning-assisted materials discovery using
  failed experiments}.
\newblock \emph{\bibinfo{journal}{Nature}} \textbf{\bibinfo{volume}{533}},
  \bibinfo{pages}{73--76} (\bibinfo{year}{2016}).

\bibitem{saal2020}
\bibinfo{author}{Saal, J.~E.}, \bibinfo{author}{Oliynyk, A.~O.} \&
  \bibinfo{author}{Meredig, B.}
\newblock \bibinfo{title}{Machine learning in materials discovery: Confirmed
  predictions and their underlying approaches}.
\newblock \emph{\bibinfo{journal}{Annual Review of Materials Research}}
  \textbf{\bibinfo{volume}{50}}, \bibinfo{pages}{49--69}
  (\bibinfo{year}{2020}).

\bibitem{vasudevan2021}
\bibinfo{author}{Vasudevan, R.}, \bibinfo{author}{Pilania, G.} \&
  \bibinfo{author}{Balachandran, P.~V.}
\newblock \bibinfo{title}{{Machine learning for materials design and
  discovery}}.
\newblock \emph{\bibinfo{journal}{Journal of Applied Physics}}
  \textbf{\bibinfo{volume}{129}}, \bibinfo{pages}{070401}
  (\bibinfo{year}{2021}).

\bibitem{jihengfang2022}
\bibinfo{author}{Fang, J.} \emph{et~al.}
\newblock \bibinfo{title}{Machine learning accelerates the materials
  discovery}.
\newblock \emph{\bibinfo{journal}{Materials Today Communications}}
  \textbf{\bibinfo{volume}{33}}, \bibinfo{pages}{104900}
  (\bibinfo{year}{2022}).

\bibitem{gubernatis2018}
\bibinfo{author}{Gubernatis, J.~E.} \& \bibinfo{author}{Lookman, T.}
\newblock \bibinfo{title}{Machine learning in materials design and discovery:
  Examples from the present and suggestions for the future}.
\newblock \emph{\bibinfo{journal}{Phys. Rev. Mater.}}
  \textbf{\bibinfo{volume}{2}}, \bibinfo{pages}{120301} (\bibinfo{year}{2018}).

\bibitem{ioannis2024}
\bibinfo{author}{Papadimitriou, I.}, \bibinfo{author}{Gialampoukidis, I.},
  \bibinfo{author}{Vrochidis, S.} \& \bibinfo{author}{Kompatsiaris, I.}
\newblock \bibinfo{title}{Ai methods in materials design, discovery and
  manufacturing: A review}.
\newblock \emph{\bibinfo{journal}{Computational Materials Science}}
  \textbf{\bibinfo{volume}{235}}, \bibinfo{pages}{112793}
  (\bibinfo{year}{2024}).

\bibitem{quanzhou2018}
\bibinfo{author}{Zhou, Q.} \emph{et~al.}
\newblock \bibinfo{title}{Learning atoms for materials discovery}.
\newblock \emph{\bibinfo{journal}{Proceedings of the National Academy of
  Sciences}} \textbf{\bibinfo{volume}{115}}, \bibinfo{pages}{E6411--E6417}
  (\bibinfo{year}{2018}).

\bibitem{peterson2021}
\bibinfo{author}{Peterson, G. G.~C.} \& \bibinfo{author}{Brgoch, J.}
\newblock \bibinfo{title}{Materials discovery through machine learning
  formation energy}.
\newblock \emph{\bibinfo{journal}{Journal of Physics: Energy}}
  \textbf{\bibinfo{volume}{3}}, \bibinfo{pages}{022002} (\bibinfo{year}{2021}).

\bibitem{tabor2018}
\bibinfo{author}{Tabor, D.} \emph{et~al.}
\newblock \bibinfo{title}{Accelerating the discovery of materials for clean
  energy in the era of smart automation}.
\newblock \emph{\bibinfo{journal}{Nature Reviews Materials}}
  \textbf{\bibinfo{volume}{3}}, \bibinfo{pages}{1} (\bibinfo{year}{2018}).

\bibitem{merchant2023}
\bibinfo{author}{Merchant, A.} \emph{et~al.}
\newblock \bibinfo{title}{Scaling deep learning for materials discovery}.
\newblock \emph{\bibinfo{journal}{Nature}} \textbf{\bibinfo{volume}{624}},
  \bibinfo{pages}{1--6} (\bibinfo{year}{2023}).

\bibitem{yunxingzuo2021}
\bibinfo{author}{Zuo, Y.} \emph{et~al.}
\newblock \bibinfo{title}{Accelerating materials discovery with bayesian
  optimization and graph deep learning}.
\newblock \emph{\bibinfo{journal}{Materials Today}}
  \textbf{\bibinfo{volume}{51}}, \bibinfo{pages}{126--135}
  (\bibinfo{year}{2021}).

\bibitem{song2021}
\bibinfo{author}{Song, Y.}, \bibinfo{author}{Siriwardane, E. M.~D.},
  \bibinfo{author}{Zhao, Y.} \& \bibinfo{author}{Hu, J.}
\newblock \bibinfo{title}{Computational discovery of new 2d materials using
  deep learning generative models}.
\newblock \emph{\bibinfo{journal}{ACS Applied Materials \& Interfaces}}
  \textbf{\bibinfo{volume}{13}}, \bibinfo{pages}{53303--53313}
  (\bibinfo{year}{2021}).

\bibitem{lyngby2022}
\bibinfo{author}{Lyngby, P.} \& \bibinfo{author}{Thygesen, K.~S.}
\newblock \bibinfo{title}{Data-driven discovery of 2d materials by deep
  generative models}.
\newblock \emph{\bibinfo{journal}{npj Computational Materials}}
  \textbf{\bibinfo{volume}{8}} (\bibinfo{year}{2022}).

\bibitem{ozonder-graphene-abs}
\bibinfo{author}{\"{O}z\"{o}nder, {\c{S}}.}, \bibinfo{author}{\"{U}nl\"{u},
  C.}, \bibinfo{author}{G\"{u}lery\"{u}z, C.} \& \bibinfo{author}{Trabzon, L.}
\newblock \bibinfo{title}{Doped graphene quantum dots uv–vis absorption
  spectrum: A high-throughput tddft study}.
\newblock \emph{\bibinfo{journal}{ACS Omega}} \textbf{\bibinfo{volume}{8}},
  \bibinfo{pages}{2112--2118} (\bibinfo{year}{2023}).

\bibitem{koppens2014}
\bibinfo{author}{Koppens, F.} \emph{et~al.}
\newblock \bibinfo{title}{Photodetectors based on graphene, other
  two-dimensional materials and hybrid systems}.
\newblock \emph{\bibinfo{journal}{Nature nanotechnology}}
  \textbf{\bibinfo{volume}{9}}, \bibinfo{pages}{780--793}
  (\bibinfo{year}{2014}).

\bibitem{yuxin2011}
\bibinfo{author}{Liu, Y.}, \bibinfo{author}{Dong, X.} \& \bibinfo{author}{Chen,
  P.}
\newblock \bibinfo{title}{Cheminform abstract: Biological and chemical sensors
  based on graphene materials}.
\newblock \emph{\bibinfo{journal}{Chemical Society reviews}}
  \textbf{\bibinfo{volume}{41}}, \bibinfo{pages}{2283--307}
  (\bibinfo{year}{2011}).

\bibitem{sarmento2018}
\bibinfo{author}{Lin, J.}, \bibinfo{author}{Huang, Y.} \&
  \bibinfo{author}{Huang, P.}
\newblock \bibinfo{title}{Chapter 9 - graphene-based nanomaterials in
  bioimaging}.
\newblock In \bibinfo{editor}{Sarmento, B.} \& \bibinfo{editor}{{das Neves},
  J.} (eds.) \emph{\bibinfo{booktitle}{Biomedical Applications of
  Functionalized Nanomaterials}}, Micro and Nano Technologies,
  \bibinfo{pages}{247--287} (\bibinfo{publisher}{Elsevier},
  \bibinfo{year}{2018}).

\bibitem{chung2021}
\bibinfo{author}{Chung, S.}, \bibinfo{author}{Revia, R.~A.} \&
  \bibinfo{author}{Zhang, M.}
\newblock \bibinfo{title}{Graphene quantum dots and their applications in
  bioimaging, biosensing, and therapy}.
\newblock \emph{\bibinfo{journal}{Advanced Materials}}
  \textbf{\bibinfo{volume}{33}}, \bibinfo{pages}{1904362}
  (\bibinfo{year}{2021}).

\bibitem{miao2012}
\bibinfo{author}{Miao, X.} \emph{et~al.}
\newblock \bibinfo{title}{High efficiency graphene solar cells by chemical
  doping}.
\newblock \emph{\bibinfo{journal}{Nano Letters}} \textbf{\bibinfo{volume}{12}},
  \bibinfo{pages}{2745--2750} (\bibinfo{year}{2012}).

\bibitem{mahmoudi2018}
\bibinfo{author}{Mahmoudi, T.}, \bibinfo{author}{Wang, Y.} \&
  \bibinfo{author}{Hahn, Y.-B.}
\newblock \bibinfo{title}{Graphene and its derivatives for solar cells
  application}.
\newblock \emph{\bibinfo{journal}{Nano Energy}} \textbf{\bibinfo{volume}{47}},
  \bibinfo{pages}{51--65} (\bibinfo{year}{2018}).

\bibitem{pang2018}
\bibinfo{author}{Pang, Y.} \emph{et~al.}
\newblock \bibinfo{title}{Facile preparation of n-doped graphene quantum dots
  as quickly-dried fluorescent ink for anti-counterfeiting}.
\newblock \emph{\bibinfo{journal}{New Journal of Chemistry}}
  \textbf{\bibinfo{volume}{42}} (\bibinfo{year}{2018}).

\bibitem{hanyung2021}
\bibinfo{author}{Jung, H.} \& \bibinfo{author}{Lee, H.}
\newblock \bibinfo{title}{Semi-transparent reduced graphene oxide temperature
  sensor on organic light-emitting diodes for fingerprint liveness detection of
  smartphone authentication}.
\newblock \emph{\bibinfo{journal}{Sensors and Actuators A: Physical}}
  \textbf{\bibinfo{volume}{331}}, \bibinfo{pages}{112876}
  (\bibinfo{year}{2021}).

\bibitem{liu2018}
\bibinfo{author}{Liu, J.} \emph{et~al.}
\newblock \bibinfo{title}{Repeated photoporation with graphene quantum dots
  enables homogeneous labeling of live cells with extrinsic markers for
  fluorescence microscopy}.
\newblock \emph{\bibinfo{journal}{Light: Science \& Applications}}
  \textbf{\bibinfo{volume}{7}} (\bibinfo{year}{2018}).

\bibitem{pirone2021}
\bibinfo{author}{Pirone, D.} \emph{et~al.}
\newblock \bibinfo{title}{Three-dimensional quantitative intracellular
  visualization of graphene oxide nanoparticles by tomographic flow cytometry}.
\newblock \emph{\bibinfo{journal}{Nano Letters}} \textbf{\bibinfo{volume}{21}},
  \bibinfo{pages}{5958--5966} (\bibinfo{year}{2021}).

\bibitem{mitra2017}
\bibinfo{author}{Mitra, R.} \& \bibinfo{author}{Saha, A.}
\newblock \bibinfo{title}{Reduced graphene oxide based “turn-on”
  fluorescence sensor for highly reproducible and sensitive detection of small
  organic pollutants}.
\newblock \emph{\bibinfo{journal}{ACS Sustainable Chemistry \& Engineering}}
  \textbf{\bibinfo{volume}{5}}, \bibinfo{pages}{604--615}
  (\bibinfo{year}{2017}).

\bibitem{bingyang2021}
\bibinfo{author}{Li, B.} \emph{et~al.}
\newblock \bibinfo{title}{Review of performance improvement strategies for
  doped graphene quantum dots for fluorescence-based sensing}.
\newblock \emph{\bibinfo{journal}{Synthetic Metals}}
  \textbf{\bibinfo{volume}{276}}, \bibinfo{pages}{116758}
  (\bibinfo{year}{2021}).

\bibitem{griffithsrr2019}
\bibinfo{author}{Griffiths, R.-R.} \& \bibinfo{author}{Hernández-Lobato,
  J.~M.}
\newblock \bibinfo{title}{Constrained bayesian optimization for automatic
  chemical design}  (\bibinfo{year}{2019}).
\newblock \eprint{1709.05501}.

\bibitem{pinheiro2020}
\bibinfo{author}{Pinheiro, G.~A.} \emph{et~al.}
\newblock \bibinfo{title}{Machine learning prediction of nine molecular
  properties based on the smiles representation of the qm9 quantum-chemistry
  dataset}.
\newblock \emph{\bibinfo{journal}{The Journal of Physical Chemistry A}}
  \textbf{\bibinfo{volume}{124}}, \bibinfo{pages}{9854--9866}
  (\bibinfo{year}{2020}).

\bibitem{honda2019}
\bibinfo{author}{Honda, S.}, \bibinfo{author}{Shi, S.} \&
  \bibinfo{author}{Ueda, H.~R.}
\newblock \bibinfo{title}{Smiles transformer: Pre-trained molecular fingerprint
  for low data drug discovery}  (\bibinfo{year}{2019}).
\newblock \eprint{1911.04738}.

\bibitem{aruspous2020}
\bibinfo{author}{Arús-Pous, J.} \emph{et~al.}
\newblock \bibinfo{title}{Smiles-based deep generative scaffold decorator for
  de-novo drug design}.
\newblock \emph{\bibinfo{journal}{Journal of Cheminformatics}}
  \textbf{\bibinfo{volume}{12}} (\bibinfo{year}{2020}).

\bibitem{wangsheng2019}
\bibinfo{author}{Wang, S.}, \bibinfo{author}{Guo, Y.}, \bibinfo{author}{Wang,
  Y.}, \bibinfo{author}{Sun, H.} \& \bibinfo{author}{Huang, J.}
\newblock \bibinfo{title}{Smiles-bert: Large scale unsupervised pre-training
  for molecular property prediction}.
\newblock In \emph{\bibinfo{booktitle}{Proceedings of the 10th ACM
  International Conference on Bioinformatics, Computational Biology and Health
  Informatics}}, BCB '19, \bibinfo{pages}{429–436}
  (\bibinfo{publisher}{Association for Computing Machinery},
  \bibinfo{address}{New York, NY, USA}, \bibinfo{year}{2019}).

\bibitem{moriconi2020}
\bibinfo{author}{Moriconi, R.}, \bibinfo{author}{Deisenroth, M.~P.} \&
  \bibinfo{author}{Kumar, K. S.~S.}
\newblock \bibinfo{title}{High-dimensional bayesian optimization using
  low-dimensional feature spaces}  (\bibinfo{year}{2020}).
\newblock \eprint{1902.10675}.

\bibitem{packwood}
\bibinfo{author}{Packwood, D.}
\newblock \emph{\bibinfo{title}{{Bayesian Optimization for Materials Science}}}
  (\bibinfo{publisher}{Springer Singapore}, \bibinfo{year}{2017}).

\bibitem{ZUO2021126}
\bibinfo{author}{Zuo, Y.} \emph{et~al.}
\newblock \bibinfo{title}{Accelerating materials discovery with bayesian
  optimization and graph deep learning}.
\newblock \emph{\bibinfo{journal}{Materials Today}}
  \textbf{\bibinfo{volume}{51}}, \bibinfo{pages}{126--135}
  (\bibinfo{year}{2021}).

\bibitem{WANG2022100728}
\bibinfo{author}{Wang, K.} \& \bibinfo{author}{Dowling, A.~W.}
\newblock \bibinfo{title}{Bayesian optimization for chemical products and
  functional materials}.
\newblock \emph{\bibinfo{journal}{Current Opinion in Chemical Engineering}}
  \textbf{\bibinfo{volume}{36}}, \bibinfo{pages}{100728}
  (\bibinfo{year}{2022}).

\bibitem{xus2021}
\bibinfo{author}{Xu, S.} \emph{et~al.}
\newblock \bibinfo{title}{Self-improving photosensitizer discovery system via
  bayesian search with first-principle simulations}.
\newblock \emph{\bibinfo{journal}{Journal of the American Chemical Society}}
  \textbf{\bibinfo{volume}{143}}, \bibinfo{pages}{19769--19777}
  (\bibinfo{year}{2021}).

\bibitem{pollice2021}
\bibinfo{author}{Pollice, R.} \emph{et~al.}
\newblock \bibinfo{title}{Data-driven strategies for accelerated materials
  design}.
\newblock \emph{\bibinfo{journal}{Accounts of Chemical Research}}
  \textbf{\bibinfo{volume}{54}}, \bibinfo{pages}{849--860}
  (\bibinfo{year}{2021}).

\bibitem{doan2020}
\bibinfo{author}{Doan, H.~A.} \emph{et~al.}
\newblock \bibinfo{title}{Quantum chemistry-informed active learning to
  accelerate the design and discovery of sustainable energy storage materials}.
\newblock \emph{\bibinfo{journal}{Chemistry of Materials}}
  \textbf{\bibinfo{volume}{32}}, \bibinfo{pages}{6338--6346}
  (\bibinfo{year}{2020}).

\bibitem{deshwal2021}
\bibinfo{author}{Deshwal, A.}, \bibinfo{author}{Simon, C.~M.} \&
  \bibinfo{author}{Doppa, J.~R.}
\newblock \bibinfo{title}{Bayesian optimization of nanoporous materials}.
\newblock \emph{\bibinfo{journal}{Mol. Syst. Des. Eng.}}
  \textbf{\bibinfo{volume}{6}}, \bibinfo{pages}{1066--1086}
  (\bibinfo{year}{2021}).

\bibitem{shieldsb2021}
\bibinfo{author}{Shields, B.} \emph{et~al.}
\newblock \bibinfo{title}{Bayesian reaction optimization as a tool for chemical
  synthesis}.
\newblock \emph{\bibinfo{journal}{Nature}} \textbf{\bibinfo{volume}{590}},
  \bibinfo{pages}{89--96} (\bibinfo{year}{2021}).

\bibitem{WAHAB2020609}
\bibinfo{author}{Wahab, H.} \emph{et~al.}
\newblock \bibinfo{title}{Machine-learning-assisted fabrication: Bayesian
  optimization of laser-induced graphene patterning using in-situ raman
  analysis}.
\newblock \emph{\bibinfo{journal}{Carbon}} \textbf{\bibinfo{volume}{167}},
  \bibinfo{pages}{609--619} (\bibinfo{year}{2020}).

\bibitem{liang2021}
\bibinfo{author}{Liang, Q.} \emph{et~al.}
\newblock \bibinfo{title}{Benchmarking the performance of bayesian optimization
  across multiple experimental materials science domains}
  (\bibinfo{year}{2021}).
\newblock \eprint{2106.01309}.

\bibitem{GUO2023109714}
\bibinfo{author}{Guo, J.} \emph{et~al.}
\newblock \bibinfo{title}{A random forest regression with bayesian
  optimization-based method for fatigue strength prediction of ferrous alloys}.
\newblock \emph{\bibinfo{journal}{Engineering Fracture Mechanics}}
  \textbf{\bibinfo{volume}{293}}, \bibinfo{pages}{109714}
  (\bibinfo{year}{2023}).

\bibitem{terayama2021}
\bibinfo{author}{Terayama, K.}, \bibinfo{author}{Sumita, M.},
  \bibinfo{author}{Tamura, R.} \& \bibinfo{author}{Tsuda, K.}
\newblock \bibinfo{title}{Black-box optimization for automated discovery}.
\newblock \emph{\bibinfo{journal}{Accounts of Chemical Research}}
  \textbf{\bibinfo{volume}{54}}, \bibinfo{pages}{1334--1346}
  (\bibinfo{year}{2021}).

\bibitem{Bombarelli}
\bibinfo{author}{Gómez-Bombarelli, R.} \emph{et~al.}
\newblock \bibinfo{title}{Automatic chemical design using a data-driven
  continuous representation of molecules}.
\newblock \emph{\bibinfo{journal}{ACS Central Science}}
  \textbf{\bibinfo{volume}{4}}, \bibinfo{pages}{268--276}
  (\bibinfo{year}{2018}).

\bibitem{Kiyohara}
\bibinfo{author}{Kiyohara, S.}, \bibinfo{author}{Oda, H.},
  \bibinfo{author}{Tsuda, K.} \& \bibinfo{author}{Mizoguchi, T.}
\newblock \bibinfo{title}{Acceleration of stable interface structure searching
  using a kriging approach}.
\newblock \emph{\bibinfo{journal}{Japanese Journal of Applied Physics}}
  \textbf{\bibinfo{volume}{55}}, \bibinfo{pages}{045502}
  (\bibinfo{year}{2016}).

\bibitem{Seko2015}
\bibinfo{author}{Seko, A.} \emph{et~al.}
\newblock \bibinfo{title}{Prediction of low-thermal-conductivity compounds with
  first-principles anharmonic lattice-dynamics calculations and bayesian
  optimization}.
\newblock \emph{\bibinfo{journal}{Phys. Rev. Lett.}}
  \textbf{\bibinfo{volume}{115}}, \bibinfo{pages}{205901}
  (\bibinfo{year}{2015}).

\end{thebibliography}

\end{document}